\documentclass[aps,prb,twocolumn,showpacs]{revtex4}
\usepackage{graphicx}

\begin{document}

\title{Superconductivity in Ti-doped Iron-Arsenide Compound Sr$_4$Cr$_{0.8}$Ti$_{1.2}$O$_6$Fe$_2$As$_2$ }

\author{Xiyu Zhu, Fei Han, Gang Mu, Peng Cheng, Bing Shen, Bin Zeng, and Hai-Hu Wen}\email{hhwen@aphy.iphy.ac.cn }

\affiliation{National Laboratory for Superconductivity, Institute of
Physics and Beijing National Laboratory for Condensed Matter
Physics, Chinese Academy of Sciences, P. O. Box 603, Beijing 100190,
People's Republic of China}

\begin{abstract}
Superconductivity was achieved in Ti-doped iron-arsenide compound
Sr$_4$Cr$_{0.8}$Ti$_{1.2}$O$_6$Fe$_2$As$_2$ (abbreviated as
Cr-FeAs-42622). The x-ray diffraction measurement shows that this
material has a layered structure with the space group of
\emph{P4/nmm}, and with the lattice constants a = b = 3.9003 $\AA$
and c = 15.8376 $\AA$. Clear diamagnetic signals in ac
susceptibility data and zero-resistance in resistivity data were
detected at about 6 K, confirming the occurrence of bulk
superconductivity. Meanwhile we observed a superconducting
transition in the resistive data with the onset transition
temperature at 29.2 K, which may be induced by the nonuniform
distribution of the Cr/Ti content in the FeAs-42622 phase, or due to
some other minority phase.
\end{abstract}

 \maketitle

The iron-based superconductors have formed a new family in the field
of high-$T_c$ superconductivity, since the discovery of
superconductivity at 26$\;$K in the layered quaternary compound
LaFeAsO$_{1-x}$F$_x$.\cite{Kamihara2008,WenAdvMat2008} Up to date,
the family of the iron-based superconductors has been extended
rapidly,\cite{WenEPL,XHCh,NLW,RenZA55K,CP,WangC,Rotter,CWCh,LiFeAs,LiFeAsChu,LiFeAsUK,Han,Tegel,HosonoSrFeAsF,ZhuXYEPL,ChengPEPL}
and the physical properties as well as pairing mechanism of these
superconductors have been investigated widely.\cite{NLWangSDW,
DaiPC, MuGHC, Luetkens, Buechner, ZhengGQ, CongRen,LeiShan, Chen,
Hashimoto, Malone, DingH, ZhouXJ, Hiraishi} Now it is known that for
the iron-arsenide based systems, superconductivity can be induced by
doping electrons or holes in the parent phases in various ways. By
the end of 2008, a new FeAs-based system,
Sr$_3$Sc$_2$O$_5$Fe$_2$As$_2$ (FeAs-32522), with the space group of
\emph{I4/mmm} and a rather large spacing distance between
neighboring FeAs-layers was successfully fabricated by our
group.\cite{FeAs32522} Recently, superconductivity was observed in
the iron-pnictide based system Sr$_4$Sc$_2$O$_6$Fe$_2$P$_2$
(FeP-42622) and also doped iron-arsenide compounds. \cite{FeP42622,
GFChen}

In this paper we present the structural and transport properties of
a new iron-arsenide based superconductor in the FeAs-42622 phase,
namely the Ti-doped Sr$_4$Cr$_{0.8}$Ti$_{1.2}$O$_6$Fe$_2$As$_2$. It
is found that this material also has a rather large distance between
neighboring conducting layers (FeAs-layers). Bulk superconductivity
was confirmed by the clear diamagnetic properties and zero
resistivity.

By using a two-step solid state reaction method, the polycrystalline
samples were fabricated. In the first step, SrAs and FeAs powders
were obtained by the chemical reaction method with Sr pieces, Fe
powders and As grains. Then they were mixed with Cr$_2$O$_3$ (purity
99.9\%), SrO (purity 99\%), TiO$_2$ (purity 99\%), Ti(purity 99\%)
and Fe powder (purity 99.9\%) in the formula
Sr$_4$Cr$_{0.8}$Ti$_{1.2}$O$_6$Fe$_2$As$_2$, ground and pressed into
a pellet shape. The pellets were sealed in a quartz tube with 0.2
bar of Ar gas, followed by a heat treatment at 1000 $^o$C for 40
hours. Then it was cooled down slowly to room temperature. All the
weighing, mixing, grounding and pressing procedures were finished in
a glove box under argon atmosphere with the moisture and oxygen
below 0.1 PPM.

The X-ray diffraction (XRD) patterns of our samples were carried out
by a $Mac$-$Science$ MXP18A-HF equipment with $\theta - 2\theta$
scan. The ac susceptibility of the samples were measured on the
Maglab-12T (Oxford) with an ac field of 0.1 Oe and a frequency of
333 Hz. The resistivity data were obtained using a four-probe
technique on the Quantum Design instrument physical property
measurement system (PPMS) with magnetic fields up to 9 T. The
temperature stabilization was better than 0.1\% and the resolution
of the voltmeter was better than 10 nV.

\begin{figure}
\includegraphics[width=9cm]{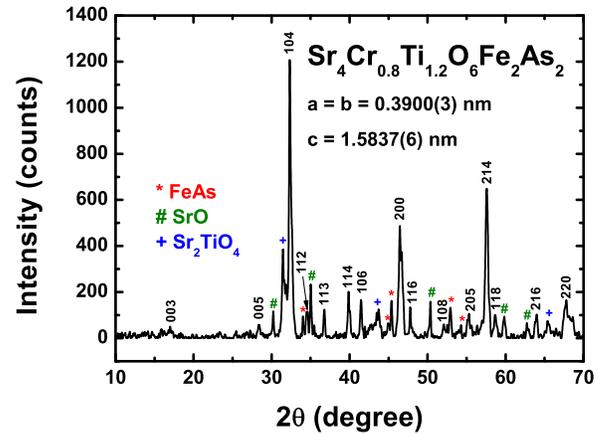}
\caption{(Color online) X-ray diffraction patterns for the sample
Sr$_4$Cr$_{0.8}$Ti$_{1.2}$O$_6$Fe$_2$As$_2$. One can see that all
the main peaks can be indexed to the structure of FeAs-42622 with
the space group of \emph{P4/nmm}. The peaks from the impurities, as
marked by the asteristics etc., are indexed to the phases FeAs, SrO
and Sr$_2$TiO$_4$.} \label{fig1}
\end{figure}

The X-ray diffraction (XRD) pattern for the sample
Sr$_4$Cr$_{0.8}$Ti$_{1.2}$O$_6$Fe$_2$As$_2$ is shown in Fig. 1. One
can see that the sample was dominated by the phase of the tetragonal
structure with the space group of \emph{P4/nmm}. Some impurity
phases were found to come from FeAs, SrO and Sr$_2$TiO$_4$. The
lattice constants of our samples were determined to be a = b =
3.9003 $\AA$ and c = 15.8376 $\AA$ from the diffraction data. We can
see that both the a-axis and b-axis lattice constants of this
material are slightly smaller than those of
(Sr$_3$Sc$_2$O$_5$)Fe$_2$As$_2$ (FeAs-32522) compound, and the
distance between neighboring conducting layers (FeAs-layers) (c =
15.8376 $\AA$) is larger than that of the FeAs-32522 phase (c/2 =
13.4 $\AA$). \cite{FeAs32522} This has become a common feature of
this system and we have argued that this feature may be intimately
related to the high-T$_c$ superconductivity in this system.

In Fig. 2 we present the temperature dependent resistivity data
under 0 T and 9 T. A metallic behavior can be seen in the
temperature region above 30 K in both fields. With the decrease of
temperature, resistivity displays a superconducting-like transition
with the onset temperature as high as 29.2 K (see inset of Fig. 2).
Then after a rather wide transition, resistivity becomes zero below
about 6 K. We think that it is possible that the superconductivity
may occur at 29.2 K for the Cr-FeAs-42622 phase. But due to the
nonuniform distribution of the doped Ti content in the sample, the
superconducting transition is broad. However, we could not exclude
another possibility that the superconductivity at 29.2 K is derived
from other minority phase. A magnetic field of 9 T didn't move the
onset transition point significantly but seemed to broaden the
transition. This suggests a rather high upper critical field for the
superconducting transition at 29.2 K. The inset of Fig. 2 shows the
enlarged view of the resistivity curve in low temperature region
under zero field.

\begin{figure}
\includegraphics[width=9cm]{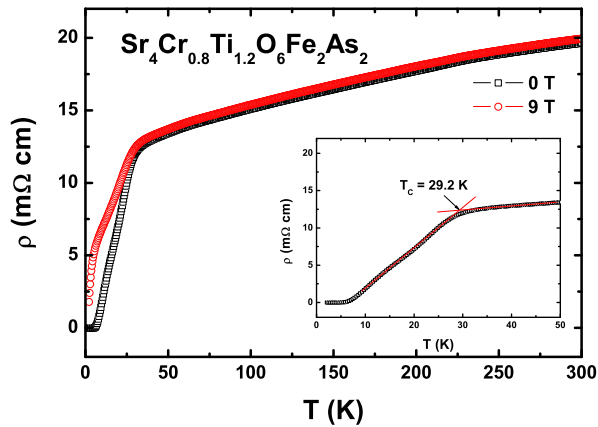}
\caption{(Color online) Temperature dependence of resistivity for
our sample Sr$_4$Cr$_{0.8}$Ti$_{1.2}$O$_6$Fe$_2$As$_2$ under two
different fields 0 T and 9 T is shown in the main frame. The inset
presents the enlarged view in the low temperature region under zero
field. One can see a superconducting-like transition with the onset
temperature of 29.2 K. Zero-resistance was observed below about 6
K.} \label{fig2}
\end{figure}

To further confirm the presence of bulk superconductivity in our
sample, we also measured the magnetization of this sample using the
ac susceptibility method. Temperature dependence of ac
susceptibility measured with $H_{ac}$ = 0.1 Oe and $f$ = 333 Hz was
shown in Fig. 3. We observed a clear diamagnetic signal in low
temperature. A rough estimate on the diamagnetic signal at about 2 K
tells that the superconducting volume is beyond 50\%, which roughly
consists with the volume calculated from the XRD data.  The onset
transition temperature from the ac susceptibility data was
determined to be about 6 K. This temperature is in good agreement on
the zero-resistance temperature obtained from the resistivity data
shown in Fig. 2. Therefore the diamagnetic signal in our sample
confirms the bulk superconductivity.

\begin{figure}
\includegraphics[width=9cm]{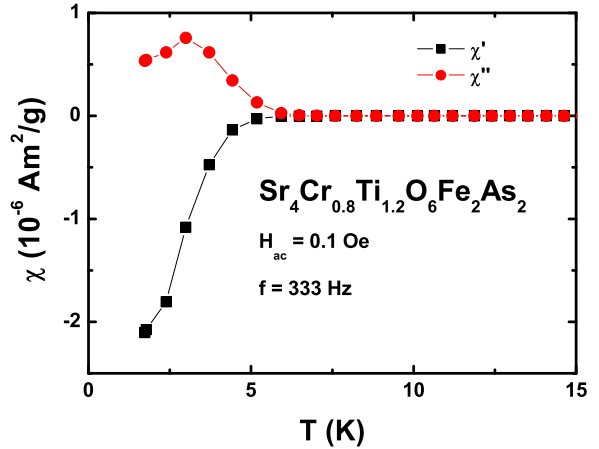}
\caption{(Color online) Temperature dependence of the ac
susceptibility measured with $H_{ac}$ = 0.1 Oe and $f$ = 333 Hz for
the sample Sr$_4$Cr$_{0.8}$Ti$_{1.2}$O$_6$Fe$_2$As$_2$. The
magnitude of the diamagnetic signal confirms the bulk
superconductivity in this sample. The onset transition temperature
was determined to be about 6 K from this figure, giving quite good
agreement with the zero-resistance temperature obtained in Fig. 2. }
\label{fig3}
\end{figure}

In summary, bulk superconductivity was achieved by substituting
Cr$^{3+}$ with Ti$^{4+}$ in Sr$_4$Cr$_{2}$O$_6$Fe$_2$As$_2$. We
investigated the structural and transport properties on one typical
sample Sr$_4$Cr$_{0.8}$Ti$_{1.2}$O$_6$Fe$_2$As$_2$. The lattice
constants, a = b = 3.9003 $\AA$ and c = 15.8376 $\AA$ are determined
from the x-ray diffraction data. It is found that this system has a
rather large spacing distance between neighboring FeAs-layers. The
resistivity and ac susceptibility data both confirmed the occurrence
of bulk superconductivity at about 6 K in the present sample. We
also observed a superconducting transition with the onset transition
temperature as high as 29.2 K in the resistivity data. At this
moment we can't distinguish whether the superconductivity at 29.2 K
found here comes from the Ti-doped Cr-FeAs-42622 phase or from other
minority phase.

We acknowledge the help of XRD experiments from L. H. Yang and H.
Chen. This work is supported by the Natural Science Foundation of
China, the Ministry of Science and Technology of China (973 project:
2006CB01000, 2006CB921802), the Knowledge Innovation Project of
Chinese Academy of Sciences (ITSNEM).

\end{document}